# DNABERT-2: Fine-Tuning a Genomic Language Model for Colorectal Gene Enhancer Classification

Darren King, Yaser Atlasi and
Gholamreza Rafiee*

I. ABSTRACT

Gene enhancers control when and where genes switch on, yet their sequence diversity and tissue specificity make them hard to pinpoint in colorectal cancer. We take a sequence-only route and fine-tune DNABERT-2, a transformer genomic language model that uses byte-pair encoding to learn variable-length tokens from DNA. Using assays curated via the Johnston Cancer Research Centre at Queen's University Belfast, we assembled a balanced corpus of 2.34 million 1 kb enhancer sequences, applied summit-centered extraction and rigorous de-duplication including reverse-complement collapse, and split the data stratified by class. With a 4,096-term vocabulary and a 232-token context chosen empirically, the DNABERT-2-117M classifier was trained with Optuna-tuned hyperparameters and evaluated on 350,742 held-out sequences. The model reached PR-AUC 0.759, ROC-AUC 0.743, and best F1 0.704 at an optimised threshold ($\tau$ = 0.359), with recall 0.835 and precision 0.609. Against a CNN-based EnhancerNet trained on the same data, DNABERT-2 delivered stronger threshold-independent ranking and higher recall, although point accuracy was lower. To our knowledge, this is the first study to apply a second-generation genomic language model with BPE tokenisation to enhancer classification in colorectal cancer, demonstrating the feasibility of capturing tumour-associated regulatory signals directly from DNA sequence alone. Overall, our results show that transformer-based genomic models can move beyond motif-level encodings toward holistic classification of regulatory elements, offering a novel path for cancer genomics. Next steps will focus on improving precision, exploring hybrid CNN-transformer designs, and validating across independent datasets to strengthen real-world utility.

*Index Terms*—Gene Enhancer, Genomic Language Model, DNABERT-2, BPE Tokenisation, Tumour Classification

II. INTRODUCTION

The Human Genome Project revealed approximately 20,000 protein-coding genes within the human genome, representing less than 2% of its 3 billion base pairs. The remaining 98% of non-coding DNA was once considered 'junk' but is now recognised as vital to regulating when, where and to what extent genes are expressed [1]. Amongst the most important of these regulatory elements are gene enhancers; these are short non-coding DNA sequences that act as molecular switches to increase transcription of their target genes. Enhancers govern processes as fundamental as cell identity, cell development, cell differentiation and cellular response to environmental stimuli [2]. Despite their central role, enhancers are difficult to identify and characterise. They lack consistent sequence motifs, can reside far from the genes they regulate and exhibit cell-type and state-specific activity [2]. Although genomic assays such as ChIP-seq and ATAC-seq have made strides in enhancer discovery, sequence-based prediction of enhancer activity remains a challenging task. This is particularly pertinent in cancer where aberrant enhancer activation can drive oncogene expression and tumour progression [3].

In parallel with advances in high-throughput sequencing, the last decade has witnessed a revolution in machine learning, particularly with the introduction of the Transformer architecture [4]. Transformers, originally developed for natural language processing, excel at modelling sequential data by capturing both local and long-range dependencies through self-attention. Their success has catalysed the development of genomic language models (gLMs) which treat DNA as a language and learn contextual representations of nucleotide sequences from massive corpora. Amongst these, DNABERT demonstrated that pre-trained transformers can achieve state-of-the-art results in DNA classification tasks by tokenising sequences into fixed k-mers [5]. However, fixed-length tokenisation is computationally costly and inflexible.

DNABERT-2 addresses these limitations by introducing byte-pair encoding (BPE) tokenisation which learns variable length nucleotide 'subwords' that better capture both motif fragments and higher-order patterns [6]. Coupled with architectural optimisations such as ALiBi positional biases and efficient attention mechanisms DNABERT-2 represents a second generation of gLMs that balances scalability with accuracy. This positions it as a promising candidate for addressing the problem of tissue-specific enhancer classification.

The present study investigates whether DNABERT-2 can accurately differentiate between normal and tumour-associated enhancers in colorectal tissue. Specifically, the aim is to fine-tune DNABERT-2 on a large corpus of enhancer sequences curated from colorectal and normal tissue samples and to benchmark its performance against previous work that employed an EnhancerNet architecture on the same dataset [7]. By focusing solely on sequence data, this project evaluates the extent to which modern transformer-based tokenisation and representation learning can capture biologically meaningful enhancer features without relying on additional epigenomic or expression data.

Darren King and Gholamreza Rafiee are with the School of Electronics, Electrical Engineering and Computer Science (EEECS), Yaser Atlasi is with The Johnston Cancer Research Centre at Queen's University Belfast (QUB), Email: {DKing11, Y.Atlasi, G.Rafiee}@qub.ac.uk

The scope of this study is intentionally constrained to binary classification of colorectal enhancer sequences, excluding multi-omic integration and cross-tissue generalisation. Nevertheless, it provides an opportunity to assess the viability of BPE tokenisation for enhancer prediction and compare transformer-based methods against a CNN-based baseline. The research objectives are: 1. To preprocess and curate a balanced dataset of normal and tumour enhancer sequences from colorectal tissue. 2. To fine-tune DNABERT-2 using BPE tokenisation for binary classification. 3. To evaluate performance using PR-AUC, ROC-AUC and F1 metrics including threshold-optimised F1. 4. To compare results against prior EnhancerNet work to contextualise improvements or limitations.

The central research question guiding this project is: *Can DNABERT-2, with its BPE-based tokenisation and transformer architecture, achieve superior or complimentary performance to CNN-based models for colorectal enhancer classification using DNA sequence data alone?*

## III. LITERATURE REVIEW

High-throughput sequencing technologies, collectively referred to as next-generation sequencing (NGS) have enabled comprehensive and rapid genome profiling, revolutionising genomic research. In parallel, advances in deep learning have produced powerful frameworks capable of modelling complex, non-linear patterns within vast biological datasets. Over the past decade, the convergence of these two developments has positioned genomics at a pivotal moment; a data-rich discipline now empowered by computational models with the scope and capacity to extract meaningful biological insights at unprecedented scale.

Against this backdrop, the 2017 seminal paper 'Attention is All You Need'[4] marked a transformative shift in the field of Deep Learning, fundamentally redefining the architecture of neural networks through the introduction of the Transformer model. The transformer is a general-purpose architecture in that it was not designed for any one task, but rather to learn from sequences of data using attention [4]. This novel mechanism termed 'self-attention', enables neural networks to weigh the significance of different data elements directly in parallel, rather than sequentially as in traditional recurrent neural networks (RNNs) or convolutional neural networks (CNNs). By eschewing recurrence and convolutions in favour of attention-only mechanisms, the Transformer has achieved unprecedented efficiency and scalability in sequence modelling, enabling greater parallelisation and accelerating training on large datasets. The immediate impact of the Transformer architecture was most pronounced in Natural Language Processing (NLP), demonstrated by substantial improvements in translation tasks, language modelling and contextual understanding. However, its conceptual innovation rapidly transcended NLP, catalysing advancements across diverse domains including computer vision [8], protein folding [9] and genomic sequence analysis [5].

In the context of gene enhancer prediction, the Transformer's capability to capture intricate long-range dependencies and sequence-contextual interaction, without losing efficiency or interpretability, positions it uniquely as an ideal architecture. It's powerful self-attention mechanism facilitates enhanced pattern recognition within DNA sequences, enabling the sophisticated modelling of regulatory elements such as enhancers, which typically involve complex interactions distributed across extensive genomic regions. Inspired by breakthroughs utilising the Transformer in NLP, researchers developed genomic language models (gLMs) – large transformer-based models pre-trained on DNA sequences. These models treat the genome as a language and learn contextual representations of nucleotide sequences by training on vast amounts of unlabelled genomic data with self-supervised objectives like masked language modelling [10]. The idea is analogous to how BERT or GPT learn representations of text, except here the "text" is genomic DNA (with an alphabet of A, C, G, T). By learning the "syntax" and "semantics" of DNA, gLMs aim to capture complex patterns that underlie biological function. This approach offers a compelling way to leverage the massive volume of genomic sequences available. Several influential gLMs have driven the field forward in the past few years. The following provides a succinct genealogy of these key gLMs:

*DNABERT (2021)* - Bidirectional Encoder Representations from Transformers for DNA [5]. DNABERT was one of the first large DNA language models. It adapted the BERT architecture to genomic sequences. DNABERT was pre-trained on the human reference genome (hg38) using a masked language modelling objective after tokenising DNA into fixed-length k-mers (e.g. 6-mers) as "words". DNABERT demonstrated that a single pre-trained model can achieve state-of-the-art performance on multiple DNA sequence classification tasks. In other words, DNABERT learned a generalisable DNA language representation allowing for transfer learning across tasks. DNABERT effectively captured global and contextual patterns in DNA, something evidenced by its strong performance and the observation that it 'correctly captures hidden syntax, grammar and semantics' of genomic sequences[5]. This opened the door to treating genomic analysis as an NLP problem.

*LOGO (2022)* – Language of Genome[11]. Whilst many gLMs grew larger, LOGO took a more lightweight approach. LOGO used an ALBERT-based transformer with only 2 self-attention layers (approximately 1 million parameters) combined with a convolutional module. LOGO was pre-trained (masked LM) on the human genome (hg19) using mixed k-mer tokens (3-6mers). Despite its small size LOGO achieved surprisingly strong results; after fine-tuning it improved promoter identification by 15% and enhancer-promoter interaction prediction by 4.5% over previous methods[11]. LOGO thus proved that efficient architectures can still capture useful genomic features, especially when coupled with convolution to inject local sequence motif awareness.

*Nucleotide Transformer (2024)* [12] – This model massively scaled gLMs into the billion-parameter arena. The Nucleotide transformer (NT) is a collection of BERT-style transformer models ranging from 50 million up to 2.5 billion parameters. These models were pre-trained on a corpus of 3,202 human genomes (taken from the 1000 Genome Project) and 850 diverse non-human genomes in addition to the reference human genome. NT primarily uses non-overlapping 6-mer tokenisation and a 6000 bp input window for training (DNABERT was limited to 512 bp). NT established a new state-of-the-art in genomic foundation models through massive scale and cross-species training.

*DNABERT-2 (2024)*[6] - An improved 'second-generation' gLM that builds on DNABERT's successes whilst addressing its limitations. DNABERT-2 made two key technical changes. Firstly, it employs a new tokeniser. Instead of using fixed k-mers as utilised in DNABERT, DNABERT-2 employs Byte-Pair Encoding (BPE) to adaptively learn a vocabulary of nucleotide sub-sequences (See: III. B – Tokenisation) This reduces the huge token set and redundancy that come with fixed k-mers. Secondly, its architecture has been optimised for efficiency by tackling the input length limit of transformers. The original DNABERT had to split sequences into circa 512 bp chunks due to the quadratic attention cost and position embeddings that don't extrapolate beyond that length. DNABERT-2 integrates FlashAttention and uses ALiBi (Attention with Linear Biases) positional encoding, which allows the model to naturally generalise to sequences longer than those seen in training. In pre-training DNABERT-2 achieved similar accuracy to the giant Nucleotide Transformers using ~21x fewer parameters and ~92x less GPU time [6]. In head-to-head comparisons, it outperformed the original DNABERT on 23 of 28 benchmark tasks despite being 3x more parameter efficient. The real takeaway from DNABERT-2 is that smart engineering can yield efficient yet powerful gLMs, rather than just brute-force scaling.

The progression from DNABERT through to DNABERT-2 illustrates the rapid evolution of gLMs with each generation addressing limitations of either tokenisation, efficiency and/or scalability. These models provide a foundation upon which to build enhancer classification/prediction pipelines. In this study, DNABERT-2 was selected as the model of choice for fine-tuning for colorectal enhancer classification. The following section details the preprocessing of normal and colorectal tumour gene enhancer sequences establishing the dataset upon which subsequent modelling and analysis were conducted.

## IV. METHODS

### A. Data Processing

The dataset employed in this study was originally derived from colorectal cancer and normal tissue samples provided by the Atlasi Lab at Queen's University Belfast. This data was partially processed in a previous MSc project [7] where the initial raw enhancer annotations were supplied in the form of .bed files corresponding to 125 colorectal cancer samples and 204 normal samples with each sample containing between 30 and 79,430 annotated enhancer regions, amounting to a total of 6,834,783 enhancer sequences. The earlier work applied an initial filtering criterion whereby samples with fewer than 5,000 enhancers were excluded. Importantly, rather than balancing the dataset at the level of samples, class balance was achieved by ensuring an equal number of enhancer sequences per class.

Following refinement, the prior study produced two consolidated datasets: 1. Normal tissue enhancers: 20 samples comprising 1,375,056 enhancer sequences. 2. Tumour tissue enhancers: 81 samples comprising 1,375,063 enhancer sequences. These refined datasets, provided as two separate .bed files, served as the starting point for the present analysis and will be referred to as the raw genomic data. The processing steps described in the subsequent sections therefore build directly upon this previously curated and balanced dataset, ensuring continuity with prior work while extending the analysis through additional preprocessing and modelling steps unique to the present study. The raw genomic data comprised ENCODE narrowPeak BED files representing enhancer intervals from both normal and tumour tissue. These files were systematically transformed into a clean, balanced dataset of labelled DNA sequences suitable for downstream modelling with DNABERT-2. *BED File Parsing:* Custom parsers were implemented to load and standardise the BED files into Pandas DataFrames. Exploratory analysis showed that the length distributions were broadly consistent across classes with $95^{th}$ percentile peak lengths of 961 bp (tumour) and 992 bp (normal) which supported subsequent sequence length choices. However, systematic differences between the classes were noted; tumour peaks exhibited higher statistical significance (mean -log10 p-value = 48.3 vs 34.0 in normal samples) however DNABERT is blind to peak-calling statistics as it only sees the sequence itself. The tumour samples also had greater source diversity (81 tumour sources versus 20 normal sources). As the samples were generated by the same lab, using the same assay protocol, same sequencing platform and the same data processing pipeline the source of the data, by itself, does not encode a confounder into the analysis so does not need to be corrected for any further. Similarly, provenance analysis also revealed differences in chromosomal representation (80 versus 71 unique contigs), raising the potential for dataset leakage and class imbalance if not controlled.

*Fixed window extraction:* A 1 kb window size was selected for sequence extraction. This decision balanced biological and computational considerations: enhancers typically average 800bp in sequence length [13] and a 1 kb context ensures inclusion of both the enhancer core and surrounding flanking regions whilst allowing the data to remain tractable for transformer architectures. Sequences were extracted from the hg19 reference genome using pyfaidx with summit-aware centring. Non-canonical contigs were excluded retaining only chromosomes 1-22, X, Y and M. Edge effects were addressed by symmetric padding with ambiguous bases (N) thereby preserving a fixed input length.

*Deduplication:* Given the scale of the dataset, deduplication was essential to prevent over-representation and label leakage. Three layers of redundancy were addressed; 1. Within class duplicates were removed. 2.Cross-class collisions were removed symmetrically.3. Reverse complement equivalence was enforced as enhancers are un-stranded.

*Ambiguity*: A negligible fraction (<0.001%) of sequences contained ambiguous bases ('N'). These were dropped to avoid introducing noise or requiring additional model logic for ambiguous symbols. *Class Balance*: Final counts were tested for balance using chi-squared and binomial proportion tests. Although both tests rejected perfect 50/50 balance due to the extremely large sample size, the effect size was negligible (tumour: normal ratio = 1.04). The dataset was therefore deemed balanced in practice for downstream tasks. (Normal tissue enhancers: 1,146,321, Tumour tissue enhancers: 1,191,958, Total: 2,338,279)

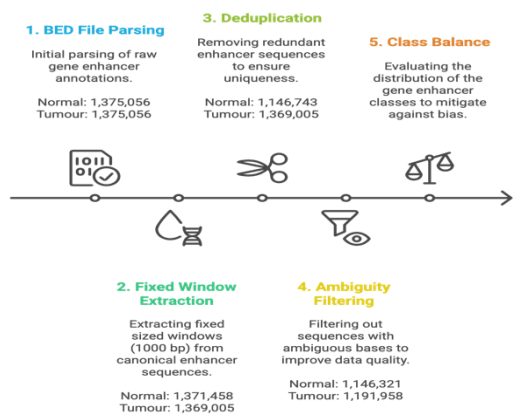

*Figure 1: Data preprocessing steps for DNABERT-2 Modelling*

*B. Tokenisation*

Following preprocessing, the curated enhancer sequences were prepared for input into the DNABERT-2 model. This required converting raw DNA sequences (strings) into numerical representations that the transformer architecture can process. The procedure involved dataset partitioning, conversion to Hugging Face Dataset objects and the application of DNABERT-2's Byte Pair Encoding (BPE)-based tokeniser. *Dataset Splitting*: The final cleaned dataset comprised 2.34 million sequences, each of fixed 1kb length with binary labels (0=normal enhancer, 1= tumour enhancer). The dataset was stratified by class and split into training (72.5%), validation (12.75%) and test (15%) sets corresponding to 1,689,406 training samples, 298,131 validation samples and 350,742 test samples. Stratified sampling ensured that both classes were proportionally represented in each split.

*DNABERT-2 Tokeniser:* Unlike the original DNABERT, which used fixed-length k-mers, DNABERT-2 employs a BPE-based subword tokeniser. BPE is a data compression algorithm that builds a fixed-sized vocabulary of variable-length tokens by iteratively merging the most frequent co-occurring segments in the corpus [6]. The process starts with an initial vocabulary of all unique characters (A,C,G,T) and then iteratively adds the most frequent multi-base segments as new words until the target vocabulary is reached. In DNABERT-2 a vocabulary size of 4096 tokens is used as it offers the best balance between model performance and computational efficiency. BPE allows the tokeniser to reduce genomic sequences to fewer tokens (roughly 4-5x compression compared to raw bases), but the compression is uneven; repetitive regions are tokenised very compactly, while irregular or low-complexity sequences produce longer tokenised outputs. This variability directly impacts the transformer's context window. To manage this the tokeniser requires setting the max_length parameter (the token length after BPE compression), which defines the maximum number of tokens per sequence after BPE. If max_length is set too small sequences will be truncated leading to information loss. If max_length is set too large, excessive padding tokens ([PAD]) are introduced inflating computation cost without contributing meaningful signal. Thus, selecting an optimal max_length was critical.

To empirically determine an appropriate context length a random sample of 100,000 sequences was tokenised without truncation. The resulting length distribution showed a median (P50) of 204 tokens, with the 90$^{th}$, 95$^{th}$ and 99$^{th}$ percentiles at 217, 220 and 225 tokens respectively. Visualisation of this distribution (Figure 2) demonstrated that almost all sequences fell below 232 tokens (P99 rounded to nearest 8 for tensor core acceleration) suggesting that truncation at this threshold would affect only a negligible fraction of samples.

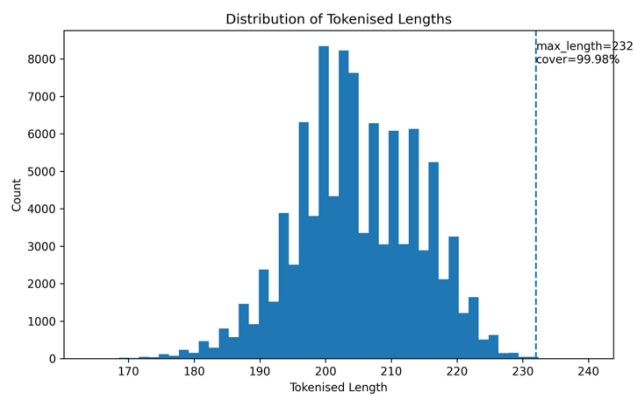

*Figure 2: Distribution of tokenised lengths of random sample of 100,000 sequences*

*Max Length Selection*: DNABERT-2-117M supports a maximum positional embedding length of 512 tokens which comfortably exceeds the observed distribution. Based on this, a final max_length of 232 tokens was selected. This configuration balances computational efficiency with biological signal preservation while aligning with hardware optimisations.

*C. Model Architecture*

The following details the architecture of DNABERT-2-117M, a Transformer-based neural network designed for sequence classification. The model, instantiated as BertForSequenceClassification, comprises a BERT encoder followed by a lightweight classification head.

*Input representation*: DNA strings are initially tokenised using a domain-specific learned vocabulary of size 4096 as discussed above. Each token is then mapped to a 768-dimensional vector via a learnable embedding matrix. (Note: the model also includes segment (token-type) embeddings to mark paired inputs when used. For single sequences, as in the case of this project, these default to a single segment). The summed embeddings are normalised with LayerNorm (epsilon=1e-12) and regularised by dropout (optimised during hyperparameter tuning). Notably, unlike vanilla BERT, DNABERT-2 does not use learned positional embedding vectors; positional information is injected later inside the attention mechanism (as detailed below).

*Encoder (Transformer stack):* The core of the model is a 12-layer BERT-style encoder. Each layer contains 2 sub-blocks, a multi-head self-attention sub-block and a feed-forward sub block each with residual connections and LayerNorm applied in a 'post-LN' pattern. *Multi-head Self-attention Sub-block*.

*Attention Type:* DNABERT-2 utilises an unpadding implementation. This removes padded tokens before computing attention and re-pads afterwards, optimising computation by avoiding unnecessary calculations on padding tokens. While this was intended to be paired with a FlashAttention kernel, this feature was disabled due to unresolved environmental dependency compatibility issues and the standard PyTorch attention routine as implemented in Hugging Face's BertSelfAttention was used. *Projections*: Queries, keys and values are generated through a fused linear projection (768 → 2304). The resulting 2304 features are then split into Q, K, V. The model employs 12 attention heads, each with a dimension of 64 (768/12). The attended output is then projected back using a linear transformation (768 → 768), followed by dropout (optimised during hyperparameter tuning), residual addition and LayerNorm.

*Positional Information:* Instead of relying on learned position embeddings, DNABERT-2 applies positional biases directly to the attention scores using ALiBi-style linear biases. This approach enables position awareness without the need for a separate position vector table. *Feed-forward (MLP) Sub-block.* DNABERT-2 replaces the standard BERT MLP with a Gated Linear Unit (GLU) MLP. A first linear layer (768 → 6144) produces 2 x 3072 channels which are split into a gate and a value stream. The gate stream is passed through GLU and multiplied elementwise with the value stream. A projection layer (3072 → 768) reduces the representation back to the model width which is then followed by dropout (optimised during hyperparameter tuning), residual addition and LayerNorm. *Pooling*: After the 12$^{th}$ layer, the model applies a pooling operation. This selects the hidden state of the first token (CLS position) and passes it through a linear layer (768 → 768) with a tanh activation. This process produces a single [batch, 768] sequence representation.

*Task Head (Classification)*: The pooled vector is regularised by dropout (optimised during hyperparameter tuning) and fed to a final linear classifier (768 → 2) yielding two logits for binary classification. In summary, the DNABERT-2-117M model features a hidden size of 768, 12 layers, 12 attention heads (each with a dimension of 64) and a GLU MLP with an internal size of 3072 (implemented via a 6144-wide gated expansion). The total parameter count is approximately 117 million.

*D. Hyperparameter Optimisation*

To maximise the predictive performance whilst maintaining computational efficiency, hyperparameter optimisation (HPO) was conducted on the DNABERT-2 model using the Optuna framework. This procedure systematically searched the parameter space to identify configurations yielding the best trade-off between accuracy, precision and recall with particular emphasis on the F1 score.

The tokenised datasets were used as inputs, with smaller subsets drawn to make HPO computationally tractable. Specifically, 100,000 training sequences and 40,000 validation sequences were sampled from the full dataset. This approach leverages the principle that relative performance between trials is more important that absolute values. Once optimal hyperparameters are identified the final model can be retrained on the full dataset.

A binary classification head was configured on DNABERT-2. The following model specific dropout parameters were tuned: hidden_dropout_prob (feed-forward layers), attention_probs_dropout_prob (attention weights), classifier_dropout (final classification head). In addition to these, the hyperparameter space included:
- Batching parameters: per-device batch size (256-768) with gradient accumulation tuned to achieve effective batch sizes of 1024-4096.
- Optimiser and learning schedule: AdamW optimiser (fused or standard); learning rate (5e-6 to 5e-4), weight decay (1e-6 to 5e-2), warmup ratio (0.04-0.12) and scheduler type (linear or cosine).
- Stability parameters: gradient clipping (max_grad_norm 0.5-2.0), label smoothing (0-0.10), Adam epsilon (1e-9 to 1e-7) and beta2 (0.98-0.9995)

As the dataset was balanced between normal and tumour enhancers, the F1 score was chosen as the primary metric. However, rather than fixing the classification threshold at 0.5, the optimal threshold was identified per trial using a threshold agnostic score (PR-AUC), sweeping through all probabilities and selecting the point that maximised F1.

To avoid excessive memory usage when evaluating large validation sets a class-based ComputeMetrics object was implemented. This design accumulated probabilities and labels across batches and computed final metrics only once per epoch enabling accurate global metrics whilst remaining compatible with Hugging Face's batch_eval_metrics=True mode. The final metrics were: 1. Precision-Recall AUC (threshold-free). 2. Maximum F1 score across thresholds. 3. Optimal threshold, precision and recall. 4. F1 at naïve 0.5 threshold (baseline comparator)

*Optuna Optimisation Strategy:* Optuna's Tree-structured Parzen Estimator (TPE) sampler with multivariate sampling was employed, coupled with a Median Pruner to terminate underperforming trials early. Each trial was limited to 1000 training steps to control computational cost with evaluation every 200 steps. A total of 50 trials were executed. Of these 20 completed, 30 were pruned and none failed. Analysis showed rapid convergence within the first 10-15 trials after which improvements were incremental. The HPO process identified a configuration that maximised predictive performance whilst maintaining computational efficiency. By combining threshold-swept F1 scoring with Optuna's efficient search strategy, the pipeline converged on a stable and generalisable set of hyperparameters. These parameters were subsequently fixed for the final training runs on the full dataset.

*E. Model Training*

The DNABERT-2-117M model was trained for binary classification with a 2-label head for normal versus tumour enhancer discrimination using the Hugging Face trainer API. Training employed the optimal hyperparameters identified during hyperparameter optimisation above, ensuring that the final configuration was both empirically validated and computationally efficient. Training was performed on an NVIDIA B200 GPU leveraging mixed precision (bf16) and Tensor Core acceleration (tf32) to optimise throughput. *Optimiser, schedule and regularisation:* Training used the AdamW optimiser (PyTorch variant) with hyperparameters imported from the best Optuna trial: learning rate $9.02 \times 10^{-6}$, weight decay $3.8 \times 10^{-6}$, $\varepsilon = 1.37 \times 10^{-9}$, $\beta_2 = 0.993$. A cosine learning-rate schedule with warmup_ratio=0.066 provided a short ramp-up followed by smooth decay. Regularisation comprised label smoothing (0.026), gradient clipping (max_grad_norm=1.43) and model level dropout tuned during HPO (hidden_dropout_prob 0.072, attention_probs_dropout_prob 0.020, classifier_dropout 0.066), which were applied to the model configuration prior to training. *Batching & Epochs:* A per-device batch size of 384 combined with gradient accumulation yielded an effective batch size of 4096. The model was trained for five epochs with evaluation conducted at the end of each epoch. Checkpointing was enabled on an epoch basis and the best checkpoint was automatically restored according to the validation PR-AUC score. *Data collation and padding efficiency:* To minimise padding overhead for variable-length BPE token sequences, batches were dynamically padded with DataCollatorWithPadding (padding to multiple of 8 for hardware efficiency) and grouped by sequence length. DataLoader settings were tuned for throughput – 8 workers, pinned memory, persistent workers and a prefetch factor of 2. *Metrics & Threshold Optimisation:* During training, validation was run via a custom, streaming ComputeMetrics object that (i) converts logits to class-1 probabilities, (ii) computes PR-AUC (threshold-free), and (iii) performs a post-hoc threshold sweep to report the single best F1 alongside F1 at the naïve 0.5 threshold. Metrics were accumulated in small CPU buffers to avoid retaining full predictions in GPU memory. A lightweight custom callback cached the best validation threshold and wrote it to disk at training end for consistent downstream use.

## V. EXPERIMENTATION & RESULTS

*Hardware & Software Environment:* The early-stage data preprocessing was conducted locally on a MacBook Pro (Apple M2 Max, 16 cores: 12 performance + 4 efficiency, 64 GB RAM). Hyperparameter optimisation, model training and evaluation were conducted on a RunPod VM: Ubuntu 24.04.2 LTS, NVIDIA B200 GPU (180GB VRAM) and 28 vCPU. The use of a RunPod VM allowed for easy use of Jupyter Lab, providing an interactive, browser-based environment that streamlined code execution, experiment tracking and visualisation while ensuring reproducibility and efficient utilisation of high-performance GPU resources. This cloud-based setup proved both simpler and more effective than employing the QUB Kelvin cluster by avoiding the overhead of queue management and environment configuration whilst offering on-demand scalability and flexibility which was better suited to the iterative nature of this project. Naturally, these advantages came with the trade-off of added expense compared to institutional cluster usage. *Training Dynamics:* Training was conducted for 5 epochs and showed stable convergence with the raw and exponentially smoothed training loss curves demonstrating a monotonic decline across steps and no evidence of late-epoch divergence (Figure 3). The learning rate schedule exhibited the expected cosine decay peaking early in epoch 1 and decaying smoothly thereafter, supporting stable optimisation (Figure 4).

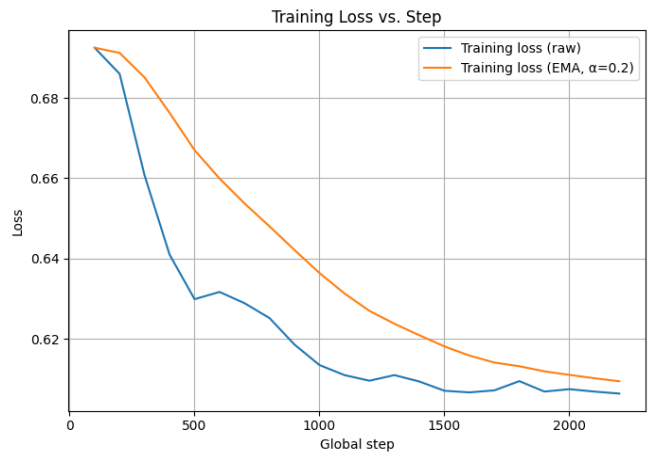

*Figure 3: Training Loss Curves*

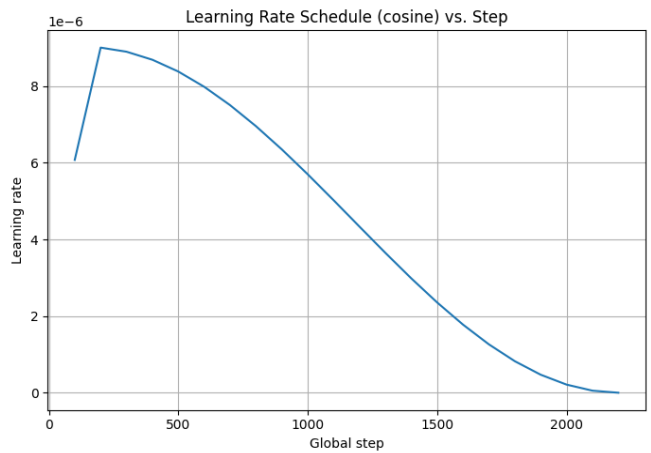

*Figure 4: Learning Rate Schedule Curve*

Validation metrics corroborated these dynamics. Evaluation loss decreased steadily before plateauing at epoch 4 (Figure 5), validation PR-AUC increased consistently before plateauing by epoch 4 (Figure 6). Effectively the model converged stably within 5 epochs with the PR-AUC plateauing at approximately 0.758 and validation loss stabilising from epoch 4 onwards.

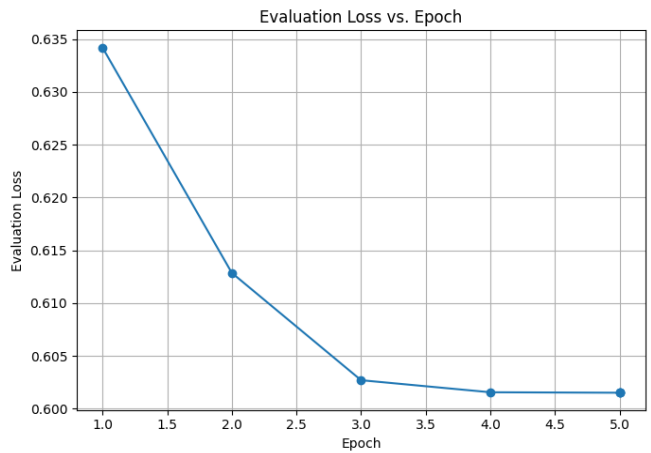

*Figure 5: Evaluation Loss Curve*

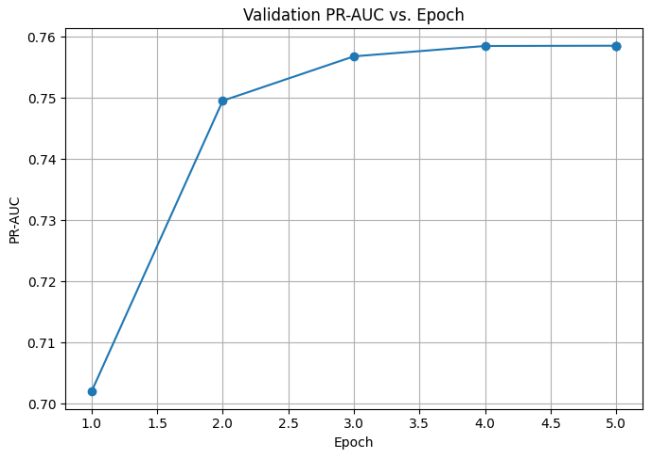

*Figure 6: Validation PR-AUC Curve*

*Test Set Performance:* The best threshold (τ=0.359) was carried forward unchanged for evaluation on the test set thereby avoiding threshold overfitting. Evaluation on the test dataset, which comprised 350,742 DNA sequences yielded the following summary statistics (Table 1). The optimal threshold for maximising the F1-score in this study was 0.359, yielding an F1 of 0.759 compared with 0.657 at the default 0.5 threshold. This demonstrates that the model benefits from threshold tuning with a lower threshold improving the balance between precision and recall. The precision recall curve confirmed this consistent ranking quality, showing that the best-threshold operating point outperformed the naïve 0.5 threshold, particularly in terms of recall. The ROC curve reported an AUC of 0.743 and the PR-AUC was 0.759 indicating broadly comparable performance across both metrics and underscoring the model's capacity to achieve useful separability and ranking quality.

*Table 1: Summary Statistics on Test Set*

| Precision Recall – AUC | 0.759 |
| --- | --- |
| ROC – AUC | 0.743 |
| Best-F1 | 0.704 |
| Precision | 0.609 |
| Recall | 0.835 |
| F1 at naïve 0.5 threshold | 0.657 |

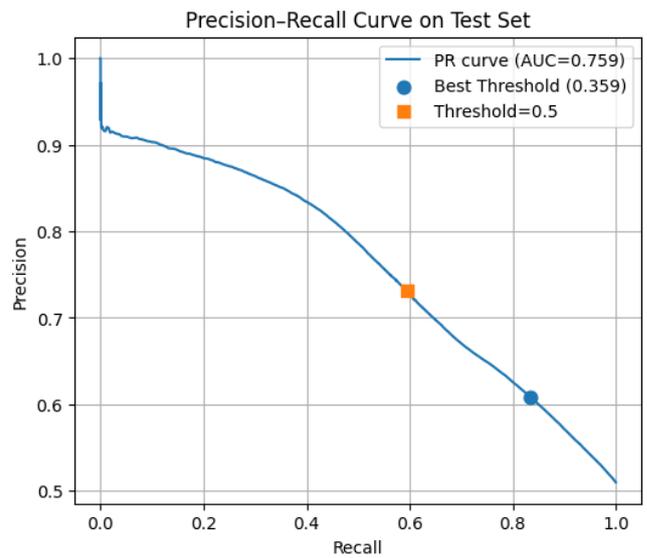

*Figure 7: Precision-recall Curve (Test Set)*

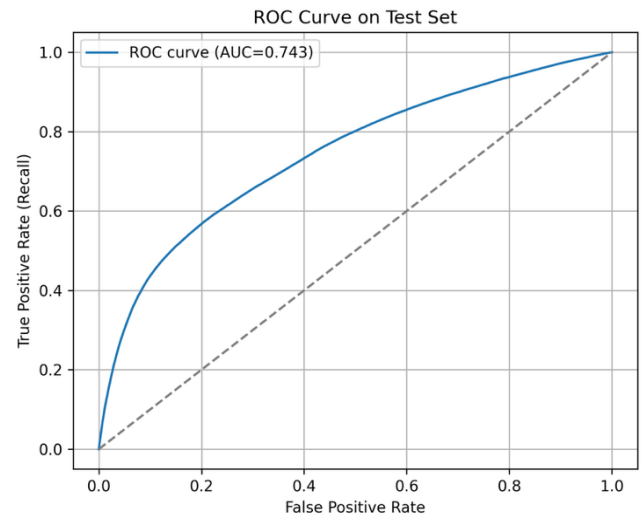

*Figure 8: ROC Curve (Test Set)*

Confusion matrices (Figure 9 & 10) illustrate the effect of threshold choice on classification performance. At the naïve 0.5 threshold precision was relatively high (0.731) but recall was suppressed (0.594) reflecting a bias towards false negatives. In contrast at the optimised threshold (τ=0.359) recall improved substantially to 0.835, with a more balanced precision-recall trade-off that yielded a higher F1 score. Notably precision was suppressed from 0.731 to 0.641 in this balancing.

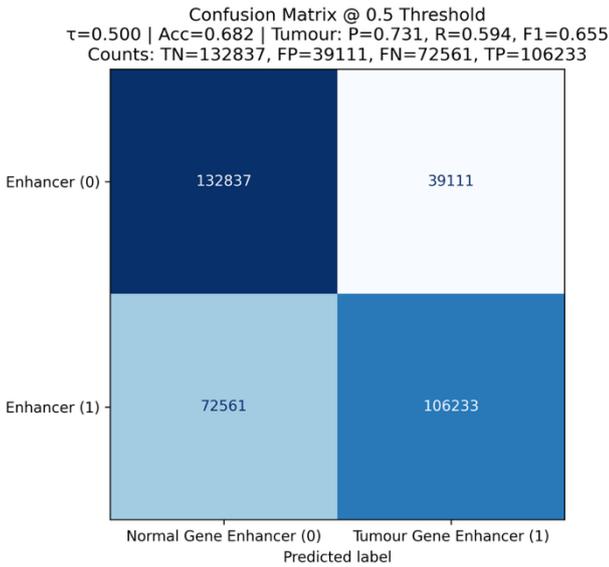

*Figure 9: Confusion Matrix at Naive Threshold*

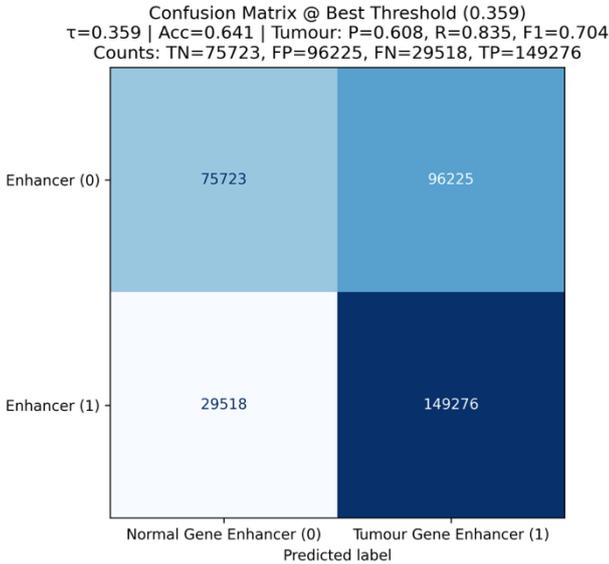

*Figure 10: Confusion Matrix at Best Threshold*

*Comparative Analysis of Test Set Performance:* The results obtained in this study using DNABERT-2 can be contrasted with those reported by a previous MSc project [7] which applied one-hot encoding and POCD-ND k-mer encodings in combination with a version of the EnhancerNet architecture on the *same dataset*. That earlier work investigated k-mer sizes of 4, 5 and 6 achieving classification accuracies of 0.70 for POCD-ND and 0.72 for one-hot encoding at k=6.

*Table 2: EnhancerNet versus DNABERT-2*

|  | **EnhancerNet** | **DNABERT-2** |
| --- | --- | --- |
| **Primary Reporting** | Accuracy, Recall, Specificity | PR-AUC ROC-AUC F1 |
| **PR-AUC** | - | 0.758 (τ=0.359) |
| **ROC-AUC** | - | 0.743 (τ=0.359) |
| **F1** | - | 0.704 (τ=0.359) |
| **Precision** | - | 0.609 (τ=0.359) |
| **Recall** | 0.72 (k=4) 0.69 (k=5) 0.72 (k=6) 0.70 (one-hot encoding) | 0.835 (τ=0.359) 0.594 (τ=0.5) |
| **Specificity** | 0.67 (k=4) 0.69 (k=5) 0.68 (k=6) 0.73 (one-hot encoding) | 0.440 (τ=0.359) 0.773 (τ=0.5) |
| **Accuracy** | 0.69 (k=4) 0.69 (k=5) 0.70 (k=6) 0.72 (one-hot encoding) | 0.641 (τ=0.359) 0.682 (τ=0.5) |

It is important to note at the outset that direct comparison between the DNABERT-2 and EnhancerNet results is challenging as the 2 studies prioritise different primary evaluation metrics. The EnhancerNet study reports point accuracy, recall and specificity, whereas the DNABERT-2 study is evaluated using threshold-free ranking metrics. At the optimised threshold of 0.359, DNABERT-2 achieved an overall accuracy of 64.1% which is significantly lower than the point accuracy of 72% reported for EnhancerNet. However, point accuracy reflects performance at a single decision threshold and does not capture the model's ability to rank tumour enhancers above normal ones across the entire probability spectrum. In contrast, the PR-AUC of 0.758 reported here demonstrates that DNABERT-2 achieves relatively strong global discriminative performance independent of any single cut-point.

Two key comparisons emerge. *1. Overall performance*: EnhancerNet with one-hot encoding achieved 0.72 accuracy while DNABERT-2 achieved a Best-F1 score of 0.704 at the optimised threshold. DNABERT-2's accuracy at this threshold was lower than EnhancerNet's but its threshold-free PR-AUC provides a more comprehensive indication of discriminative ability. These differences highlight how model assessment depends heavily on the chosen evaluation metric: accuracy provides a single-point estimate, whereas PR-AUC reflects performance across all thresholds. *2. Recall-specificity trade off:* Both studies noted tumorous sequences were easier to identify than normal ones. In EnhancerNet this was reflected in the slightly higher recall values relative to specificity. Similarly, DNABERT-2 achieved recall of 0.835 considerably higher than its precision of 0.609 showing the same tumour-sensitivity bias. Importantly, the ability of DNABERT-2 to tune decision thresholds post-hoc allowed recovery of balanced operating points, something that the k-mer based approach could not as easily exploit. Taken together, these comparisons suggest that DNABERT-2 offers a more flexible framework for enhancer classification with BPE tokenisation and transformer attention allowing the model to capture both motif-like features and longer-range sequence context. This appears to improve ranking ability and recall relative to the earlier k-mer based approach. However, the comparison also highlights limitations; while DNABERT-2 shows superior ranking and sensitivity, its precision remains modest at best. Future work should therefore benchmark both models under unified evaluation metrics to better contextualise these results.

## VI. Discussion

The primary objective of this study was to evaluate the application of transformer-based genomic language models for the classification of normal and tumour associated gene enhancers and to assess whether DNABERT-2, which employs byte-pair coding (BPE) could provide advantage over more traditional feature encoding methods. The findings demonstrate that DNABERT-2 achieves moderately robust discriminative performance on this dataset, with PR-AUC = 0.758 and Best-F1 = 0.704 at an optimised threshold. These values reflect consistent and stable convergence across training, validation and test sets indicating that the model is capable of reliably ranking tumour enhancers above normal ones. A key observation is DNABERT-2's strong recall (0.735), which underscores the model's ability to detect tumour-associated gene enhancers. This is particularly relevant in the biological context where false negatives (i.e. failing to identify tumour enhancers) may be more costly than false positives. At the same time, the model's precision remains modest at best (0.61) highlighting a tendency toward overcalling tumour sequences. This mirrors the tumour-sensitivity bias observed in the prior EnhancerNet based work, although DNABERT-2 achieved higher recall and stronger threshold-independent performance as captured by the PR-AUC. The implications of these findings extend beyond this dataset. Demonstrating that DNABERT-2 generalises well to enhancer classification strengthens the case for applying transformer-based models to broader regulatory genomic tasks. The ability to rank enhancers effectively suggests potential downstream use in prioritising candidate regions for functional validation where recall is critical. Moreover, the relatively well-calibrated probabilities observed on the test set mean that DNABERT-2 outputs can be meaningfully integrated into probabilistic decision-making frameworks without requiring extensive post-hoc calibration.

## VII. Conclusion & Future Work

Despite the promising results, several limitations must be acknowledged. *Domain knowledge in preprocessing*: The data preprocessing stage arguably required a deeper knowledge of genetics than this researcher possessed. Whilst every effort was made to mitigate bias through summit-centred extraction, deduplication and filtering, it is possible that subtle genomic considerations were not fully accounted for. This project is primarily rooted in Artificial Intelligence, however, as is common in interdisciplinary research, expertise in the secondary domain is often critical to ensuring robust preprocessing. Time constraints limited the ability to fully deepen this knowledge, and it cannot be claimed with absolute confidence that all potential sources of biological bias were eliminated. Further work with a biologist should be considered in ensuring effective preprocessing of the data. *Scale of computation*: The dataset comprised millions of enhancer sequences making preprocessing, hyperparameter optimisation and fine tuning computationally intensive. Even with access to an NVIDIA B200 GPU training was protracted: hyperparameter optimisation often spanned several days and full fine tuning required lengthy runtimes. Whilst this is not a limitation of the research per se (large-scale computation is the reality of working with deep learning and large language models) it is a practical challenge. The scale of computation, and the degree of resource required was underestimated at the outset of the project and this inevitably constrained the number of experimental iterations feasible within the project's timeframe. *Model and dependency challenges*: Working with DNABERT-2 presented technical obstacles. FlashAttention could not be enabled due to dependency issues, forcing its deactivation and likely reducing efficiency. Additionally, the Hugging Face implementation of DNABERT-2 required modification of the classification-head source code to prevent the unintended return of hidden states alongside logits during prediction which had initially caused severe memory inefficiency and repeated OOM issues. In terms of future work, an important next step would be to directly compare the performance of the current DNABERT-2 model to the EnhancerNet model. However, to ensure a fair and meaningful comparison both models should be benchmarked under unified evaluation metrics to better contextualise their results. A key avenue for future research would be to explore hybrid architectures that take the essence of the EnhancerNet study and this study by combining convolutional neural networks (CNNs) with transformers in a conformer-style design. CNNs excel at capturing local motif-level features (e.g. transcription factor binding sites) whereas transformers are more effective at modelling long-range dependencies across enhancer regions. A combined model may therefore yield more balanced precision and recall by integrating both local and contextual information. In addition to (and not necessarily independent of) the above, efforts should be directed towards improving precision. This may be achieved through more robust data preprocessing with specialist insight or possibly through ensemble strategies. The aim would be to reduce false positives and improve the practical utility of predictions in applications further down the line.

Finally, validation on independent datasets, ideally from different tissues or experimental sources will be crucial in assessing generalisability and avoiding dataset-specific overfitting. The DNABERT-2 paper proposes just such a dataset with its Genome Understanding Evaluation Plus benchmark[6]. This could be used to further validate a fine-tuned DNABERT-2 binary classifier.

In summary, this project has demonstrated the potential of transformer-based genomic language models to advance enhancer classification, offering improvements in recall and global discriminative ability compared to earlier approaches. At the same time, it has highlighted the persistent challenges of working at the interface of AI and genomics; domain expertise, computational scale and the evolving maturity of model implementations in a relatively nascent field. Whilst DNABERT-2 represents a significant step forward, the limitations acknowledged here illustrate both the difficulty and the promise of applying Artificial Intelligence to complex biological questions. Future work that integrates hybrid architectures, broader benchmarking and cross-dataset validation holds the potential to build on these findings and contribute to more reliable and interpretable enhancer prediction models.